\documentclass{ws-p8-50x6-00}

\begin{document}

\title{``Glueballs'': results and perspectives from the Lattice}

\author{Gunnar S. Bali}

\address{Department of Physics and Astronomy, University of Glasgow,
G12 8QQ, UK\\
E-mail: g.bali@physics.gla.ac.uk}


\maketitle
\abstracts{
I review the present status of lattice calculations
of properties of ``gluon-rich'' hadrons 
and comment on future prospects, in view of
planned experiments.}
\section{Review of lattice results}
The gluons of QCD should not only manifest themselves
in deep inelastic scattering but also affect the hadron spectrum.
In the sector of pseudoscalar mesons this is indeed
the case: gluodynamics results in the axial anomaly which in
turn implies a big mass gap between the singlet $\eta$ and the
octet of SU(3) Goldstone pions.
In addition to such indirect
effects, QCD in principle offers the possibility of bound states
made entirely out of glue.

In lattice simulations of
the so-called
quenched approximation (or quenched model) to QCD, i.e.\ 
QCD without sea quarks,
a rich spectrum
of glueballs has been established in the past
decade.\cite{Bali:1993fb,Morningstar:1999rf}
The scalar ($J^{PC}=0^{++}$) turns out to be lightest
with a mass between 1.4 and 1.8~GeV, followed by a tensor
of mass between 1.9 and
2.3~GeV and a pseudoscalar that is heavier by another
150~MeV\cite{Bali:1993fb,Sexton:1995kd,Morningstar:1999rf,Lucini:2001ej,Bali:2001gf}. All but five states turn out to be heavier
than 3~GeV,\cite{Bali:1993fb,Morningstar:1999rf} overlapping
with charmonia states, a mass region that future
experiments might shed more light onto.\cite{Peters:2001xc}
The other striking features are
the somewhat
counter-intuitive spin ordering of the spectrum, e.g.\ $1, 3, 2, 0$
in the $PC=+-$ sector
but $0,2,3,1$ in the $++$ sector as well as the fact that the lightest
spin-exotic state is well above 4~GeV.

The scalar glueball is of particular
phenomenological interest.\cite{Close:2001zp}
While all raw lattice data agree with each other
within statistical errors of about 40~MeV, rather different values
are quoted in the literature:\cite{Bali:1993fb,Sexton:1995kd}
in QCD an experimental input is required to set the mass scale.
However, in the quenched model
ratios of light hadronic masses
can easily deviate from {\em real world}
experiment by as much as 10~\%.\cite{Aoki:2000yr}
Hence, to some degree the translation into physical units is a matter of
personal preference. This
uncertainty is accounted for in the mass ranges quoted above.

In {\em real QCD} with sea quarks,
it is not entirely obvious in how far e.g.\ a vector glueball
that contains $c\bar{c}$ sea quarks can be distinguished from
a $J/\psi$ that contains sea gluons: no pure
glueballs exist but then neither do
pure quark model mesons and yet
the $J/\psi$ is distinctively different from a $\phi$ that shares the
same quantum numbers. We can interpret the former as being close
to a quenched $c\bar{c}$ state and the latter as a dominantly
$s\bar{s}$ state. In QCD some almost pure glueballs might exist.
It might also be that some QCD states can be understood in terms
of mixing between glueballs and mesons of the quenched model.
In some sectors it might even happen that an interpretation in terms
of mixing breaks completely down and the gluons merely result
in extra states that are hard to distinguish from radial excitations.
So-called spin-exotic quantum numbers 
like $J^{PC}=0^{--}, 0^{+-}, 1^{-+}, 2^{+-},\cdots$
are of particular interest in the search for gluon-rich states,
i.e.\ ``glueballs'' and (quark-gluon) ``hybrid mesons'', however,
even in this sector exotic four-quark ``molecules'' and
hybrid mesons can have very similar signatures.

On the theoretical side two directions of research are being pursued:
quenched and ``un-quenched''.
The quenched model is
a natural extension of the quark model and provides
the language required to speak about mixing between quark model
states and glueballs.
In order to make the connection to phenomenology
glueballs are not enough but
the corresponding flavour singlet meson states have to be
studied too. Lattice simulations indicate
that the quenched
$s\overline{s}$ isoscalar meson is about 200~MeV lighter than
the scalar glueball.\cite{Lee:2000kv,McNeile:2001xx}
Another important question is that of molecules.
Despite of some attempts in this
direction\cite{Alford:2000mm}
this possibility
is vastly unexplored at present.
The $f_0(980), a_0(980)$ and the $f_0(400-1200)$ are widely believed to be
$K\overline{K}$ and $\pi\pi$
resonances,\cite{Pennington:1999fa,Close:2001zp} however, this view
which is important for the interpretation of the
$f_0(1370), f_0(1500)$ and $f_0(1710)$\cite{Close:2001zp}
as mixtures between a scalar glueball and the two
lightest isoscalar quark model mesons,
is not completely un-debated.\cite{Minkowski:1999hq}
Molecules might also be required to explain the
difference between the spin-exotic $1^{-+}$ mesons observed
around 1.4 and 1.6 GeV in experiment but predicted around 1.9~GeV
in lattice studies.
The next step would be to look into mixing.
In addition to the first exploratory lattice
investigation\cite{McNeile:2001xx} several models have been
proposed.\cite{Amsler:1996td,Lee:2000kv,Anisovich:2001zr}
In some references the $f_0(1500)$ receives the dominant
gluonic contribution,\cite{Amsler:1996td} in others
it is the $f_0(1710)$.\cite{Lee:2000kv}
Finally, production and
decays reveal information about the quark content of a given resonance,
provided one knows what to expect from a glueball.
Lattice methods are only of limited
use here although an exploratory study
does exist.\cite{Sexton:1995wg,Lee:2000kv}

\begin{figure}[t]
\begin{center}
\includegraphics[width=20pc]{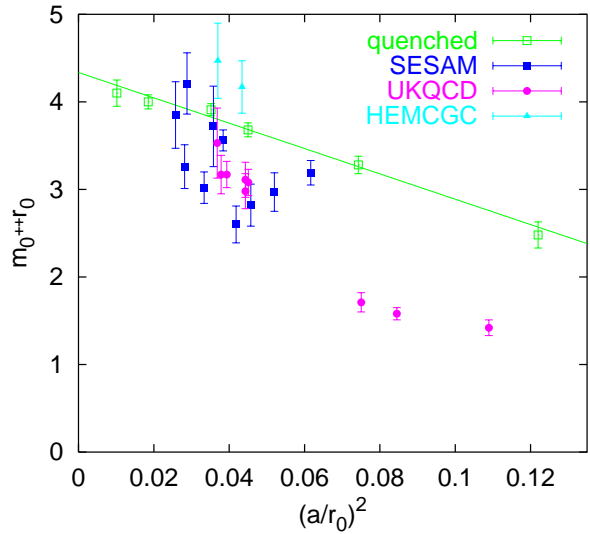}
\end{center}\vskip -.3cm
\caption{The scalar ``glueball'' with $n_f=2$
vs.\ the lattice spacing $a$ ($r_0^{-1}\approx 400$~MeV).
\label{fig:glue}}
\vskip -.8cm
\end{figure}
The second, cleanest path is to compute the spectrum of QCD
{\em as is}. One can then compare with experiment and
hopefully find agreement.
Unfortunately, the direct approach does at present
not only turn out to be prohibitively expensive computationally 
but it does not really tell us what we want to know either:
just like in real experiment we would only be able to determine masses
and, by varying the lattice volume, decay widths within a given
channel but little would be revealed about the
{\em nature} of the states. Mixing cannot be studied
because the resonances are just {\em out there}. Fortunately,
in our virtual computer experiment
we can gradually reduce the quark mass, starting
from the quenched approximation, and trace any changes,
in particular in the neighbourhood of decay or
mixing thresholds.

We are still in the position that
the combined ``world data'' on the scalar $n_f=2$ ``glueball''
fits into Fig.~\ref{fig:glue}.
The quenched
case\cite{Lucini:2001ej,Bali:1993fb} is included for reference.
The un-quenched results have been obtained by use of three
different lattice discretisations of the Dirac action:
staggered (HEMCGC\cite{Bitar:1991wr}),
Wilson (SESAM\cite{Bali:2000vr})
and clover
(UKQCD\cite{McNeile:2001xx,Hart:2001fp}).
The quarks are all
heavier than $m_s/3$, the scalar meson is still stable and
the wave function
turns out to be very close to that of the quenched
glueball.\cite{Bali:2000vr,McNeile:2001xx} Most $n_f=2$
points clearly lie below the quenched line, however, there
is certainly a slope in the results,
such that the mass in the physical $a=0$ limit appears
consistent with
the quenched result. Within the SESAM data set
there is an apparent discontinuity because different points
have been obtained at different quark masses;
the ``glueball'' becomes lighter
as the quark mass is reduced.
Whether this effect weakens as the continuum limit is approached
is a question as open as whether anything will substantially
change once the quarks have become realistically light.

\section{Outlook}
The quenched glueball spectrum is ``solved'' and first promising
$n_f=2$ results exist. Future studies of mixing in the quenched set-up
are important.
Flavour singlet mesons and molecules as well as standard
charmonium spectroscopy has been neglected in the past for various
reasons but lattice methods and
computers have sufficiently matured to allow for a fast quenched
relief.
More challenging but ultimately necessary is an analysis of
the quark mass, volume and lattice spacing dependence
with $n_f=2+1$ sea quarks.
\section*{Acknowledgments}
G.B.\ is a Heisenberg Fellow (DFG grant Ba~1564/4-1) and has received
support from PPARC grant PPA/G/O/1998/00559.

\end{document}